\documentclass[a4paper,10pt]{article}
\usepackage{amssymb,graphicx}
\makeatletter
\def\eqnarray{%
   \stepcounter{equation}%
   \def\@currentlabel{\p@equation\theequation}%
   \global\@eqnswtrue
   \m@th
   \global\@eqcnt\z@
   \tabskip\@centering
   \let\\\@eqncr
   $$\everycr{}\halign to\displaywidth\bgroup
       \hskip\@centering$\displaystyle\tabskip\z@skip{##}$\@eqnsel
      &\global\@eqcnt\@ne\hskip 0.5\arraycolsep \hfil$\displaystyle{##}$\hfil
      &\global\@eqcnt\tw@ \hskip 0.5\arraycolsep
         $\displaystyle{##}$\hfil\tabskip\@centering
      &\global\@eqcnt\thr@@ \hb@xt@\z@\bgroup\hss##\egroup
         \tabskip\z@skip
      \cr
}
\makeatother
\begin{document}

\begin{center}
{\Large\bf Reheating Temperature \\ after Inflation in String-inspired Supergravity} \\
\vspace{16pt}
Yuta Koshimizu$^{a}$, Toyokazu Fukuoka$^{a}$, Kenji Takagi$^{b}$, Hikoya Kasari$^{a}$\\
 and Mitsuo J. Hayashi$^{a}$ \\
\vspace{16pt}
$^{a}${\it Department of Physics, Tokai University, 1117, Kitakaname, Hiratsuka, 259-1292, Japan} \\
$^{b}${\it Company VSN, Minato, Shibaura, 108-0023, Japan} \\

\vskip 16pt
E-mail: mhayashi@keyaki.cc.u-tokai.ac.jp
\vskip 16pt
\today
\end{center}

\thispagestyle{empty}
\vspace{24pt}

\begin{abstract}
By using a string-inspired modular invariant supergravity model, which was proved well to explain WMAP observations appropriately, a mechanism of preheating just after the end of inflation is investigated. 
By using the canonically normalized and diagonalized scalars, the decay rates of these fields are calculated inflaton $S$ into gauge sector fields. 

The reheating temperature is estimated by both the stability condition of Boltzmann equation and the instant preheating mechanism. Both of reheating temperatures are almost the same order of magnitude $\sim O(10^{10})$ GeV. Because two mechanisms are completely independent processes, the former is caused through the inflaton decays into gauge fields and gauginos and the latter is caused by the scattering process of two inflatons into two right handed sneutrinos, which will decay into Higgs fields and other minimal SUSY standard model (MSSM) particles, we conclude that both mechanisms play essential roles in the preheating process after inflation.
\end{abstract}

\newpage
\section{Introduction}
\addtocounter{page}{-1}

Following ``Seven-Year Wilkinson Microwave Anisotropy Probe (WMAP) Observations" \cite{ref:WMAP}, the theory of inflation are proved to be the most promising theory of the early universe before the big bang.

 As a favored scenario to explain the observational data, it is customary to introduce a scalar field called inflaton.
What kind of theoretical frameworks are the most appropriate to describe the theory of particle physics, inflation and the recent observations?
It seems to require a far richer structure of contents than that of the standard theory of particles. 
 As far as the 4$D$, $N=1$ supergravity can play an elementary role in the theory of  the space-time and the particles \cite{ref:SUGRA}, it can also be essential in the theory of the early universe as an effective field theory. 
Supergravity, however, has been confronting with the difficulties, such as the $\eta$-problem and the supersymmetry breaking (SSB) mechanism has been studied by many authors \cite{ref:Nilles, ref:NoScale, ref:Witten, ref:Cvetic}.
We have investigated to prevail over these difficulties in Refs.\cite{ref:Hayashi1, ref:Hayashi2} by using the modular invariant supergravity induced from superstring \cite{ref:Ferrara}. We found that the interplay between the dilaton field $S$ and gauge-singlet scalar $Y$ could give rise to sufficient inflation. The model is free from the the $\eta$-problem and realizes appropriate amount of inflation as well as the TT angular power spectrum.  

In this letter, the preheating mechanism just after the end of inflation will be investigated. 

First we will briefly review the model and the former results \cite{ref:Hayashi1, ref:Hayashi2} as follows.
It is convenient to introduce the dilaton field $S$, a chiral superfield $Y$ and the modular field $T$. Here, all the matter fields are set to zero for simplicity.
Then, the effective No-Scale type K{\"a}hler potential and the effective superpotential that incorporate modular invariance are given by \cite{ref:Ferrara}, (see also \cite{ref:Nilles0}):
\begin{eqnarray}
K&=&-\ln \left(S+S^\ast\right)-3\ln \left(T+T^\ast-|Y|^2\right), \\
W&=&3bY^3\ln\left[c\>e^{S/3b}\>Y\eta^2(T)\right],
\end{eqnarray}
where $\eta(T)$ is the Dedekind $\eta$-function, defined by:
\begin{equation}
\eta(T)=e^{-2\pi T/24} \prod^{\infty}_{n=1}(1-e^{-2\pi nT}).
\end{equation}
The parameter $b$ and $c$ are treated as free parameters in this letter as discussed in Ref.\cite{ref:Hayashi2}.
We had found that the potential $V(S,Y)$ at $T=1$ has a stable minimum at for the values
$b = 9.4$, $c = 131$ and obtained
\begin{eqnarray}
S_{{\rm min}} = 1.51, \qquad Y_{{\rm min}} = 0.00878480,
\end{eqnarray}
where $\eta (1) = 0.768225$, $\eta^2 (1) = 0.590170$, $\eta' (1) = -0.192056$, $\eta'' (1) = -0.00925929$ are used.
The inflationary trajectory can be well approximated by
\begin{equation}
Y_{\rm min}(S) \sim 0.009268 e^{-0.035461 S},
\end{equation}
which corresponds to the trajectory of the stable minimum for both $S$ and $Y$.
The slow-roll parameters $\varepsilon_S$ and $\eta_{SS}$ satisfy the slow-roll conditions.
The number of $e$-folds $\sim 57$, by integrating from $S_{\rm end} \sim 4.19$
to $S_*\sim 11.6$, i.e. our potential can produce
a cosmologically plausible number of $e$-folds \cite{ref:WMAP}.
Here $S_*$ is the value corresponding to the scale $k_*=0.05$ Mpc$^{-1}$.
We can also compute scalar spectral index and its running that describe the scale dependence
of the spectrum of primordial density perturbation $\mathcal{P_R} = (H/\dot{S})^2 ( H/2\pi )^2$ \cite{ref:Perturbation};
these indices are defined by
\begin{eqnarray}
n_s - 1 &=& \frac{d\ln \mathcal{P_R}}{d\ln k}, \\
\alpha_s &=& \frac{dn_s}{d\ln k}.
\end{eqnarray}
These are approximated in the slow-roll paradigm as
\begin{eqnarray}
n_s (S) &\sim& 1 - 6 \varepsilon_S + 2 \eta_{SS}, \\
\alpha_s(S) &\sim& 16 \varepsilon_S \eta_{SS} - 24 \varepsilon_S^2 - 2\xi^2_{(3)},
\end{eqnarray}
where $\xi_{(3)}$ is an extra slow-roll parameter that includes the trivial third derivative of
the potential.
Substituting $S_*$ into these equations, we have $n_{s* } \sim 0.951$ and
$\alpha_{s*} \sim -2.50 \times 10^{-4}$.

Because $n_s$ is not equal to 1 and $\alpha_{s}$ is almost negligible, our model suggests
a tilted power law spectrum.
The value of $n_{s*}$ is consistent with the recent observations;
the best fit of seven-year WMAP data using the power law $\Lambda$CDM model is
$n_s \sim 0.963 \pm 0.014$ \cite{ref:WMAP}.
Finally, estimating the spectrum $\mathcal{P_R}$ in slow-roll approximation (SRA),
\begin{equation}
\mathcal{P_R}\sim\frac{1}{12\pi^2}\frac{V^3}{\partial V^2},
\end{equation}
we find $\mathcal{P_R}_* \sim 2.36\times10^{-9}$.
This result matches the measurements as well \cite{ref:WMAP, ref:Hayashi1, ref:Hayashi2}. Incidentally speaking, the energy scale $V\sim10^{-10}$ GeV is also within the constrained range obtained by Liddle and Lyth \cite{ref:Liddle}.

In order to study on the angular power spectrum, we need the tensor perturbation (the gravitational wave) spectrum which is given as follows:
\begin{equation}
\mathcal{P}_{\rm grav} = 8 \left( \frac{H}{2\pi} \right)^2 = \frac{2}{3\pi^2}V.
\end{equation}
In SRA, the spectral index of $\mathcal{P}_{\rm grav}$ is given by the slow-roll parameters
$\epsilon$ and $\eta$ as
\begin{equation}
n_{T} = -2\epsilon.
\end{equation}
Using these parameters $TT$ and $TE$ angular spectrum were well fitted to the WMAP data \cite{ref:Hayashi2}.

\section{Gravitino mass and the other mass parameters}

Now we will briefly investigate the properties of inflaton $S$, gravitino and SSB mechanism. 
First, gravitino mass is given in this case 
\begin{equation}
m_{3/2} = M_P e^{K/2} |W| = 3.16 \times 10^{12} \,\, {\rm GeV},
\end{equation}
where $\hbar = 6.58211915 \times 10^{-25}$ GeV$\cdot$sec and $M_p = 2.435327 \times 10^{18}$ GeV are used.

In our model, by expanding the potential $V$ around the minimum of $S(t)$, $Y(t)$ and
 fixed $T=1$, and by providing $S(t)$ and $Y(t)$ are real, then we obtained $S(t)$, $Y(t)$ as follows:
\begin{eqnarray}
&&S(t) = S_{{\rm min}}+\sqrt{\frac{8}{3}}\frac{\sin(m_S t)}{m_S t}, \\
&&Y(t) = \frac{1}{\eta^2 (1) e^{1/3} c} e^{-\frac{S(t)}{3b}}.
\end{eqnarray}
After scalars $S,Y,T$ are canonically normalized and the masses diagonalized \cite{ref:Endo}, \cite{ref:nakamura}, the mass eigenstates are denoted by $S',Y',T'$, then masses are calculated as 
$M_{S'}=3.97 \times 10^{12}$ GeV, $M_{Y'}=2.45 \times 10^{17}$ GeV, $M_{T'}=9.02 \times 10^{12}$ GeV, where $S'$, $Y'$, $T'$ are defined as follows: 
\begin{eqnarray}
&&S' = 3.00 \times 10^{-1} S + 1.94 \times 10^{-3} Y - 3.66 \times 10^{-1} T \\[5pt]
&&Y' = 3.82 \times 10^{-4} S + 1.22 \, Y - 2.94 \times 10^{-8} T \\[5pt]
&&T' = 1.40 \times 10^{-1} S - 7.49 \times 10^{-3} Y + 7.85 \times 10^{-1} T.
\end{eqnarray}

Supersymmetry is overwhelmingly broken by superfield $S'$, which will be shown in separate paper comparing with the fact pointed out by Nilles et al. \cite{ref:Nilles1, ref:Nilles2, ref:Nilles3}, in which the interchange of SSB fields occurs (see also Kalolosh et. al. \cite{ref:Kallosh}). 

Canonically normalized fermionic states of supersymmetric partners $\tilde{S}$, $\tilde{Y}$, $\tilde{T}$ are given by
\begin{eqnarray}
\tilde{S}' = 0.331 \tilde{S}, \qquad
\tilde{Y}' = 1.22 \tilde{Y}, \qquad
\tilde{T}' = 0.867 \tilde{T} - 7.61 \times 10^{-3} \tilde{Y},
\end{eqnarray}
and the values of them are numerically determined as
\begin{eqnarray}
m_{\tilde{S}'} = 0 \,\, {\rm GeV}, \qquad
m_{\tilde{Y}'} = 3.01 \times 10^{17} \,\, {\rm GeV}, \qquad
m_{\tilde{T}'} = 2.65 \times 10^{15} \,\, {\rm GeV}.
\end{eqnarray}
Since $\tilde{S}$ is massless and $S$ breaks supersymmetry, $\tilde{S}$ state is identified with Goldstino, which is absorbed into gravitino by super-Higgs mechanism \cite{ref:SUGRA, ref:moroi}. 

Non-thermal production of gravitinos is not generated from the inflaton (dilaton), since the inflaton mass is lighter than gravitino, but they are produced by the decay of modular field $T$ and scalar field $Y$, which processes are shown in our separate paper.

\section{Decay rate from inflaton to gauginos}

Because the canonically normalized mass eigen state inflaton $S'$ does not decay into gravitinos, $S'$ will decay directly decay into the minimal SUSY standard model (MSSM) particle or the next to minimal SUSY standard model (NMSSM) particles after the end of inflation.
As an example, the decay rate of $S'$ into gauginos is estimated in our model. 
By using the term 
${\mathcal{L}}_{gaugino}=\kappa \int d^2\theta f_{ab}(\phi)W_{\alpha}W^{\alpha},\ f_{ab}(\phi)=\phi\delta_{ab}$, 
the interaction between $S$ and gauginos $\lambda^a$'s are given as
\begin{eqnarray}
&&{\mathcal{L}}_{gaugino}
=\frac{i}{2}f^R_{ab}(\phi)\left[\lambda^a\sigma^\mu\tilde{\mathcal{D}}_\mu\bar{\lambda}^b+\bar{\lambda}^a\sigma^\mu\tilde{\mathcal{D}}_\mu\lambda^b \right]-\frac{1}{2}f^I_{ab}(\phi)\tilde{\mathcal{D}}_\mu\left[\lambda^a\sigma^{\mu}\bar{\lambda}^b\right] \nonumber \\
&&\qquad\qquad\quad -\frac{1}{4}\frac{\partial f_{ab}(\phi)}{\partial \phi}e^{K/2}G_{\phi\phi^*}D_{\phi^*}W^*\lambda^a\lambda^b+\frac{1}{4}\left(\frac{\partial f_{ab}(\phi)}{\partial \phi}\right)^*e^{K/2}G_{\phi\phi^*}D_{\phi}W\bar{\lambda}^a\bar{\lambda}^b. \label{gaugino_decay}
\end{eqnarray}
By seeing the first term of (\ref{gaugino_decay}), $\lambda^a$'s are also canonically normalized as $\lambda^a= \left<f^R_{ab}\right>^{-\frac{1}{2}}\hat{\lambda}^a$. 

The interactions come from the third and fourth terms. The terms include $e^{K/2}G^{\phi\phi^*}D_{\phi^*}W^*$, which implies the auxiliary field of $\phi$ in global SUSY theory and it is replace by $F_\phi$.

By expanding $\frac{\partial f_{ab}}{\partial\phi}F_\phi$ in the terms around the stable point, interaction terms are given as
\begin{eqnarray}
&&{\mathcal{L}}_{\rm{int}}=-\frac{1}{4\left<f_{ab}\right>}\left[\left<\frac{\partial^2 f_{ab}}{\partial\phi^2}F_\phi+\frac{\partial f_{ab}}{\partial\phi}\frac{\partial F_\phi}{\partial\phi}\right>\delta\phi+\left<\frac{\partial f_{ab}}{\partial\phi}\frac{\partial F_\phi}{\partial\phi^*}\right>\delta\phi^* \right]\lambda^a\lambda^b \nonumber \\
&&\qquad\qquad\quad-\frac{1}{4\left<f_{ab}\right>}\left[\left<\frac{\partial^2 f^*_{ab}}{\partial{\phi^*}^2}F^*_\phi+\frac{\partial f^*_{ab}}{\partial\phi^*}\frac{\partial F^*_{\phi^*}}{\partial\phi^*}\right>\delta\phi^*+\left<\frac{\partial f^*_{ab}}{\partial\phi^*}\frac{\partial F^*_{\phi^*}}{\partial\phi}\right>\delta\phi \right]\bar{\lambda}^a\bar{\lambda}^b,
\end{eqnarray}
where when $\phi = S$, $F_S$ implies the SSB scale of the model and will be estimated as $\left< S+S^* \right> \gg m_{3/2}$ since $\left< F_S \right> \sim m_{3/2}$ and $(S+S^*)$ take value about $3$ times of Planck scale. Therefore, as the first term contribute far smaller than the second and negligible, $- \left< \frac{\partial F_S}{\partial S} \right> \sim m_{3/2}$ is remained.

The derivative term by $S^*$ can be replaced by $- \left< \frac{\partial F_S}{\partial S^*} \right> \sim m_{S}$. Then the decay rate $\Gamma(\phi\to \lambda+\lambda)$ can be estimated as:
\begin{eqnarray}
\Gamma(S\to \lambda + \lambda)=\frac{3}{16\pi}\frac{\left<\alpha^i_j\right>^2}{\left<f_{ab}\right>^2}m^2_\lambda m_S\left(1+\frac{m^2_{3/2}}{m^2_S}+2\frac{m_{3/2}}{m_S}\right)\left(1-\frac{4m_\lambda^2}{m_S^2} \right)^\frac{1}{2}.
\end{eqnarray}
By using the relation $F_S \sim M_pm_{SP}$ that holds for the mass of SUSY particles,
the order of gaugino mass $m_\lambda$ will be estimated as $O(10^{7})$ GeV $\sim O(10^{8})$ GeV. Then the decay rate of inflaton to gauginos can be estimated to be $\Gamma(S\to \lambda+\lambda)\sim 3.89 \times 10^{3}$ GeV \cite{ref:Polchinski}.

\section{Reheating temperature}

Now we will calculate the reheating temperature from two methods and compare the resulting temperatures. The first method is using Boltzmann equation and the decay rate of the inflaton already calculated above. 
Another rely on the instant parametric resonance to calculate the evolution of the number density of the produced matters and obtain reheating temperature. 

First, the reheating temperature $T_R({\rm gaugino})$ is derived from Boltzmann equation by using the decay rate, is given by
\begin{eqnarray}
T_R({\rm gaugino})=\left(\frac{10}{g_*}\right)^\frac{1}{4}\sqrt{M_P~\Gamma(S\to \lambda + \lambda) },
\end{eqnarray}
and numerically given  as $T_R \sim 4.45 \times 10^{10}$ GeV, by inserting the decay rate from the canonically normalized inflaton field $S'$. 
where $g_*$ is the number of the effective degrees of freedom of MSSM, i.e.  $g_* =228.75$.

Next, we will estimate the reheating temperature $T_R$ by instant parametric resonance by introducing NMSSM superpotential with the righthanded neutrino superfields.
We should choose a model to determine the reheating temperature by inflaton decay into MSSM or NMSSM particles. We will assume the instant preheating mechanism \cite{ref:fkl}, because this method is mathematically easier to control than that of parametric resonance \cite{ref:kls}.

It is not unique to take in the ordinary particles into string-inspired modular invariant supergravity. 
We will assume that minimal K{\"a}hler potential $\sum_i \Phi_i{\Phi^*}^i\label{nK}$ is simply added to $K$ and super potential of NMSSM is added to include a term that directly coupled with the inflaton (dilaton) superfield during the oscillations of inflaton. 
We will choose it as that the righthanded neutrino superfields $N^c$ couple directly with the inflaton \cite{ref:fkn}:
\begin{eqnarray}
W_{NMSSM}=M_{R_i}N^c_iN^c_i+\lambda_iSN^c_iN^c_i+\gamma^{ij}_\nu N^c_iL_jH_u \label{NMSSM}.
\end{eqnarray}
Number density through the parametric resonance is given by \cite{ref:kls}
\begin{eqnarray}
\quad n_k^{j+1} = e^{-\pi \kappa^2} + \left( 1 + 2 e^{-\pi \kappa^2} \right) n_k^j - 2 e^{-\pi \kappa^2 /2} \sqrt{1+e^{-\pi \kappa^2}} \sqrt{n_k^j (1+n^j_k)} \sin \theta^j.
\end{eqnarray}
If we consider the instant preheating mechanism \cite{Felder:1998vq}, the number density is  simply given by the first term
\begin{eqnarray}
n_k = e^{-\pi \kappa^2} = e^{-\frac{\vphantom{l_{l_{l_a}}} \pi (k^2/a^2+M_{R_i})}{\vphantom{l^l}\lambda_i|\dot{S}_0|}}, 
\end{eqnarray}
where we have used NMSSM superpotential (\ref{NMSSM}).
By integrating this equation in $k$, we derive the number density of supersymmetric partner of the righthanded neutrinos $n_{\tilde{N}^c_i}$.
\begin{eqnarray}
n_{\tilde{N}^c_i} = \frac{1}{2\pi^2} \int_0^{\infty} dk k^2n_k
= \frac{(\lambda_i\dot{S}_0)^{3/2}}{8\pi^3} \exp \left(-\frac{\pi M^2_{R_i}}{\lambda_i|\dot{S}_0|}\right) \label{number_density}.
\end{eqnarray}
From this equation we can estimate the reheating temperature
\begin{eqnarray}
T_R ({\rm Instant}) = \left(\frac{30}{\pi^2g_*}\cdot M_{R_i} \cdot n_{\tilde{N}^c_i}\right)^{1/4}
= \left( \frac{15 M_{R_i} (\lambda_i\dot{S}_0)^{3/2}}{4\pi^5g_*} \exp \left(-\frac{\pi M^2_{R_i}}{\lambda_i|\dot{S}_0|}\right) \right)^{1/4}.
\end{eqnarray}
Since $\dot{S}_0 = 6.39 \times 10^{30}$ in our model, only free parameters in $T_R ({\rm Instant})$ are $\lambda_i$ and $M_{R_i}$. We assume here to restrict $\lambda_i=M^2_{R_i}/|\dot{S}_0|$ so as to the index of the exponent in the number density (\ref{number_density}) takes the value $O(1)$, then
\begin{eqnarray}
T_R ({\rm Instant}) = M_{R_i} \left(\frac{15}{4\pi^5g_*} e^{-\pi} \right)^{1/4},
\end{eqnarray}
$T_R ({\rm Instant})$ only depends on $M_{R_i}$. If we estimate the value of $M_{R_i} \sim m_S$ as $O(10^{12})$ GeV, $\lambda_i$ takes the value $O(10^{-6})$ and $T_R ({\rm Instant})$
becomes $O(10^{10})$ GeV, which value is similar value as $T_R({\rm gaugino})$.
Therefore we conclude that both the contribution from
inflaton decays and that from the parametric resonance of four body scattering process play equally important roles in the preheating process just after the end inflation.

Because the primordial gravitinos decay very rapidly and the reheating temperature is lower than the gravitino mass,  the effect to the standard Big Bang Nucleosynthesis (BBN) scenario \cite{ref:takahashi, ref:kawasaki-moroi, ref:Kawasaki, ref:kawasaki2, ref:riotto, ref:Buchmuller}) may be negligible in our model (see also \cite{Weinberg:1982zq, Khlopov:1984pf}).

\section{Conclusion}

We have investigated on the preheating mechanism just after the end of inflation through both the inflaton (dilaton) decay into MSSM gauge sector and the collision of two inflaton into two righthanded sneutrinos. 
We conclude that the contribution of both the inflaton decays and the parametric resonance of four body scattering process play equally important roles in the preheating process just after the end inflation.
The model we used, cleared the $\eta$-problem and appeared to predict successfully the values of observations at inflation era.
It predicted for examples, the indices $n_{s* } \sim 0.951$ and
$\alpha_{s*} \sim -2.50 \times 10^{-4}$. The value of $n_{s*}$ is consistent with the recent observations; the best fit of seven-year WMAP data using the power law $\Lambda$CDM model is
$n_s \sim 0.963 \pm 0.014$ \cite{ref:WMAP}.
The estimation of the spectrum was as $\mathcal{P_R}_* \sim 2.36\times10^{-9}$, which
result matches the measurements as well \cite{ref:WMAP, ref:Hayashi1, ref:Hayashi2}.  
 
The reheating temperature $T_R$ is estimated by assuming the instant preheating mechanism and by using a tentative model among NMSSM models \cite{ref:fkl} and comparatively the same order  with that from the decay process of inflaton into MSSM gauge sector, which value is about order $\sim O(10^{10})$ GeV.
Because the mass of gravitino is calculated as $3.16 \times 10^{12}$ GeV, it is rather heavy and may be unstable, therefore, may not be considered as the lightest supersymmetric particle (LSP) or the next lightest (NLSP) and not a dark matter candidate discussed in Refs.\cite{ref:Endo, ref:nakamura, ref:Pradler}. However the main topic of supergravity at present stage of the theory is whether the gravitino exist or not in nature despite its mass.
It is not reproduced after the reheating of the universe. The gravitinos are produced almost instantly just after the end of inflation through $Y$ and $T$, not from inflaton. Then the yield variable for gravitino may take rather large value, however the decay time appears very rapid and disappear before the BBN stage of the universe. Though the effects of this type of gravitinos in the evolution of the universe should be investigated more carefully, the topic must be remained to later works.
Therefore, we only remark here that our present model seems consistent with the present situation of observations.

On the other hand, we commented that supersymmetry is overwhelmingly broken by $F-$term of the inflaton (dilaton) superfield $S$, that may be contrary to the occurrence of the interchange of SSB fields pointed out by Nilles et al. \cite{ref:Nilles1, ref:Nilles2, ref:Nilles3}. More detailed investments will be shown in our later works. 

Though we have been exclusively restricted our attention to a model of Ref.\cite{ref:Ferrara}, the other models derived from the other type of compactification seems very interesting. Among them KKLT model \cite{ref:Linde, Endo:2005uy, Choi:2005uz, ref:Nilles4} attracts our interest, where the moduli superfield $T$ plays an essential roles. We should take all the circumstances into consideration on essential problems confronted in construction of (string-inspired modular invariant) Supergravity models.

%%%%%%bibliography%%%%%%%%%

\end{document}